\documentclass[sn-mathphys,Numbered]{sn-jnl}

\usepackage{graphicx}%
\usepackage{multirow}%
\usepackage{amsmath,amssymb,amsfonts}%
\usepackage{amsthm}%
\usepackage{mathrsfs}%
\usepackage[title]{appendix}%
\usepackage{xcolor}%
\usepackage{textcomp}%
\usepackage{manyfoot}%
\usepackage{booktabs}%
\usepackage{algorithm}%
\usepackage{algorithmicx}%
\usepackage{algpseudocode}%
\usepackage{listings}%
\usepackage{braket}
\usepackage{bm}

\begin{document}

\title[DQNN via PFE]{Distributed Quantum Neural Networks via Partitioned Features Encoding}


\author*[1]{\fnm{Yoshiaki} \sur{Kawase}}\email{ykawase@g.ecc.u-tokyo.ac.jp}

\affil*[1]{\orgdiv{Graduate School of Information Science and Technology}, \orgname{The University of Tokyo}, \orgaddress{\street{7-3-1 Hongo, Bunkyo-ku}, \city{Tokyo}, \postcode{113-8656}, \country{Japan}}}


\abstract{
Quantum neural networks are expected to be a promising application in near-term quantum computing, 
but face challenges such as vanishing gradients during optimization and limited expressibility by a limited number of qubits and shallow circuits. 
To mitigate these challenges, an approach using distributed quantum neural networks has been proposed to make a prediction by approximating outputs of a large circuit using multiple small circuits. 
However, the approximation of a large circuit requires an exponential number of small circuit evaluations. 
Here, we instead propose to distribute partitioned features over multiple small quantum neural networks 
and use the ensemble of their expectation values to generate predictions. 
To verify our distributed approach, we demonstrate ten class classification of the Semeion and MNIST handwritten digit datasets. 
The results of the Semeion dataset imply that
while our distributed approach may outperform a single quantum neural network in classification performance, 
excessive partitioning reduces performance.
Nevertheless, for the MNIST dataset, we succeeded in ten class classification with exceeding 96\% accuracy. 
Our proposed method not only achieved highly accurate predictions for a large dataset but also reduced the hardware requirements for each quantum neural network 
compared to a large single quantum neural network.
Our results highlight distributed quantum neural networks as a promising direction for practical quantum machine learning algorithms compatible with near-term quantum devices.
We hope that our approach is useful for exploring quantum machine learning applications. 
}



\maketitle

\section{Introduction}\label{sec:intro}
Quantum machine learning has emerged as a promising application for near-term quantum computers. 
Popular quantum machine learning algorithms like quantum kernel methods \cite{havlivcek2019supervised,schuld2019quantum,haug2023quantum} and quantum neural networks (QNNs) \cite{cerezo2021variational,mitarai2018quantum, farhi2018classification, schuld2014quest} have been studied. 
QNNs, particularly deep QNNs, exhibit remarkable expressibility \cite{abbas2021power,sim2019expressibility,schuld2021effect},  
but the limitations of current quantum devices, such as restricted qubit counts and constrained circuit depths, reduce the model complexity. 
In addition, QNNs suffer from optimization problems of vanishing gradients during optimization \cite{mcclean2018barren,cerezo2021cost,wang2021noise}. 

To mitigate these problems, a promising direction in QNNs research is the development of distributed algorithms across multiple quantum devices \cite{pira2023invitation}. 
An advantage of distributed QNNs is reported that a kind of distributed QNNs offers an exponential reduction in communication for inference and training compared to classical neural networks using gradient descent optimization \cite{gilboa2023exponential}. 
Additionally, a distributed approach allows us to accelerate simulation by directly partitioning a given problem for parallel computation, 
for example, by distributing data to multiple quantum circuits for digit recognition or distributing the calculation of partitioned Hamiltonians for variational quantum eigensolvers \cite{du2021accelerating}.
Another approach enables us to approximate the evaluation of outputs of a large quantum circuit 
by reconstructing it from the results of small quantum circuits \cite{marshall2023high}. 
This is achieved by a circuit cutting technique, i.e., expressing two-qubit gates as a sum of the tensor products of single-qubit unitaries \cite{bravyi2016trading}. 
This approach, however, requires an exponential number of small quantum circuit evaluations. 
Further research into distributed QNNs frameworks that fully utilize multiple quantum computers while overcoming hardware limitations is required.

In this paper, we introduce a novel approach utilizing distributed QNNs for processing separately partitioned features over multiple QNNs to avoid using circuit cutting techniques. 
Specifically, we partition the input features and separately encode them into distinct QNNs. 
We then sum the expectation values from the QNNs to make predictions. 
Unlike the recent study in Ref.~\cite{wu2022wpscalable} that used distributed QNNs via encoding significantly downscaled partitioned features for feature extraction and another QNN to perform binary classification of the MNIST dataset via encoding the extracted features,
our approach requires fewer qubits and can handle the entire MNIST dataset of $28\times28$ features and $60000$ training data. 
So, our approach is more efficient for large features and multi-class classification. 
We numerically investigate the performance of our proposed distributed QNNs approach.

First, we compared the classification performance between a single QNN and our distributed approach on the Semeion handwritten digit dataset \cite{misc_semeion_handwritten_digit_178}. 
Due to the high dimensionality of the original $16 \times 16$-dimensional data, 
which was beyond our simulation capabilities especially when simulating it with a single large QNN,
we reduced the $16 \times 16$ features to $8 \times 8$ via average pooling. 
This preprocessed data was then classified using either a single QNN or our distributed QNNs.
The results demonstrated that our distributed approach achieved higher accuracy and lower loss compared to the single QNN. 
Further, we extended our distributed method to handle the original $16 \times 16$-dimensional data, employing configurations of four and eight independent QNNs. 
While both distributed models also showed a nice performance, 
the result of eight QNNs achieved higher accuracy but at the expense of increased loss, compared to the four QNNs. 
These results imply that encoding all features into a single QNN may not be an optimal approach 
and too many partitions degrade performance. 

Furthermore, in order to validate the scalability of our distributed QNNs approach, 
we applied our distributed approach to the MNIST handwritten digit dataset \cite{lecun2010mnist}. 
This dataset consists of $60000$ training data and $10000$ test data, each of $28\times28$ size. 
By employing $14$ QNNs with our distributed approach, 
we achieved exceeding $96\%$ accuracy in ten class classifications of this large dataset.
This accomplishment is particularly notable considering the computational demand of classically simulating the multiclass classification task on MNIST with a single QNN. 

Our results highlight distributed QNNs as an effective and scalable architecture for quantum machine learning, with applicability to real-world problems. 
We anticipate our proposed method will aid future distributed QNNs research and investigations into quantum advantage by enabling experiments on large practical datasets.

\section{Results}\label{sec:results}
In this section, we present the validation of our distributed QNNs approach 
through the classification of the Semeion and MNIST handwritten digit datasets.
In the following, we briefly describe the setup of our numerical experiments.
First, we focused on the Semeion dataset\cite{misc_semeion_handwritten_digit_178},  
containing $1593$ of $16 \times 16$-dimensional data representing digits from $0$ to $9$, with each feature value assigned an integer from $0$ to $255$.

For preprocessing the Semeion dataset, 
we adopted a normalization strategy for angle encoding. 
Specifically, we normalized the data values 
between $0$ and $\pi/8$ for encoding $64$ features per QNN, 
and between $0$ and $\pi/4$ for encoding $32$ features per QNN. 
Then, we applied $2\times2$ average pooling to the normalized data 
to reduce the dimension $16 \times 16$ to $8 \times 8$ 
due to the limitation of our GPU memory capacity, especially when simulating the classification task with a single QNN. 

The architectural designs of our single QNN and distributed QNNs are illustrated in Fig.~\ref{fig:qc} and described further in Appendix~\ref{secA:model_description}. 
The primary distinction between these architectures lies in the number of qubits and encoding layers, which are adjusted according to the size of partitioned features allocated to each QNN in the distributed setup.
In our approach, we distributed the evenly partitioned features across multiple independent QNNs.
Then, we evaluated the expectation values using a set of observables $\{X_1,\ldots,X_5,Z_1,\ldots,Z_5\}$ for each QNN, 
followed by summing over $n_\text{qc}$ expectation values,
where we denote $n_\text{qc}$ as the number of QNNs. 
We applied the softmax function to this aggregate multiplied by a constant factor
and then evaluated cross entropy loss as our loss function. 
We optimized the parameters $\{\bm{\phi}_j\}_{j=1}^{n_\text{qc}}$ in QNNs 
to minimize the loss function using Adam optimizer \cite{kingma2014adam} with learning rate $0.005$. 
To perform our numerical experiments efficiently,
we utilized ``torchquantum'' \cite{hanruiwang2022quantumnas} library, 
known for efficient classical simulation of quantum machine learning. 

\subsection{Results for the Semeion dataset}
First, we focused on the classification of the $8\times8$ dimensionally reduced Semeion dataset 
using both a single QNN and two QNNs model.
For the two QNNs model, we partitioned the features so that each QNN processed four rows of the data.
The results of those models with five-fold cross-validation are shown in Table~\ref{tbl:semeion}. 
The comparative results revealed that
the two QNNs model outperformed the single QNN model in terms of accuracy and loss. 
The results underscore the potential benefits of our distributed QNNs approach over a single QNN. 

Encouraged by this result, 
we sought to examine the scalability of our approach with an increased number of partitions, 
analyzing how the performance of distributed QNNs changes with more partitions. 
Therefore, we extended our exploration to the classification of the original $16\times16$ Semeion dataset, employing with four and eight QNNs. 
In these setups, each QNN processes four or two rows of features per QNN, respectively. 
While both the four QNNs and eight QNNs model demonstrated effective performance, 
the result of the eight QNNs model in loss is inferior to the four QNNs model. 
This result implies that distributing excessively partitioned features across multiple QNNs decreases performance 
since we optimize parameters to minimize loss.

In conclusion, our results indicate that encoding all features into a single QNN is not always the best approach. Moreover, our results indicate that the distribution of excessively partitioned features across multiple QNNs leads to a decline in overall performance.

\begin{table}[ht]
\caption{The ten class classifications accuracy of the Semeion dataset with $5$-fold cross-validation.
We denote the standard deviation as stddev in the table.}
\label{tbl:semeion}
\begin{tabular*}{\textwidth}{@{\extracolsep\fill}lcccc}
\toprule%
& & & \multicolumn{2}{@{}c@{}}{Semeion data set ($8\times8$)} \\\cmidrule{4-5}%
model & n\_qubits & n\_features/1QNN & accuracy $\pm$ stddev & loss $\pm$ stddev \\
\midrule
QNN                  & $8$ & $64$ & $0.93909 \pm 0.01285$ & $1.55759 \pm 0.01290$ \\
Distributed 2QNNs    & $8$ & $32$ & $0.94787 \pm 0.01499$ & $1.54413 \pm 0.01175$ \\
\botrule \toprule%
& & & \multicolumn{2}{@{}c@{}}{Semeion data set ($16\times16$)} \\\cmidrule{4-5}%
model & n\_qubits & n\_features/1QNN & accuracy $\pm$ stddev & loss $\pm$ stddev \\
\midrule 
Distributed 4QNNs    & $8$ & $64$ & $0.94788 \pm 0.01255$ & $1.54790 \pm 0.01256$ \\
Distributed 8QNNs    & $8$ & $32$ & $0.94851 \pm 0.00790$ & $1.56486 \pm 0.01459$ \\
\botrule
\end{tabular*}
\end{table}

\subsection{Results for the MNIST dataset}
We further validated scaling to more partitions by classifying MNIST handwritten digit dataset \cite{lecun2010mnist}, 
containing $60000$ training data and $10000$ test data representing digits from $0$ to $9$. 
These samples are characterized by a higher dimensionality of $28 \times 28$, 
with feature values ranging from $0$ to $255$.
As preprocessing, 
we normalized the values between $0$ and $\pi/4$ for angle encoding on single-qubit rotation gates, 
maintaining consistency with the preprocessing methodology used in our Semeion experiments. 
We distributed the equally partitioned features across $14$ QNNs, i.e., encoding two rows of features into each QNN. 
The same set of observables was employed here as well.
From the result of this numerical experiment, as shown in Table~\ref{tbl:mnist}, 
our distributed approach achieved exceeding $96\%$ accuracy for the test data, 
demonstrating the robustness of our distributed approach against performance degradation. 
In addition, our distributed approach operates accurately at this scale, 
which is infeasible to simulate classically using a single QNN. 
This success highlights the potential of our distributed QNNs approach
to be a highly effective and scalable architecture for practical quantum machine learning.
Our findings imply that distributed QNNs could play an important role in advancing the field of quantum machine learning.

\begin{table}[ht]
\caption{The ten class classifications performance for the MNIST dataset: 
the model with a star mark(*) uses a mini-batch to reduce the required GPU memory while training.}
\label{tbl:mnist}
\begin{tabular*}{\textwidth}{@{\extracolsep\fill}lcccc}
\toprule
& & & \multicolumn{2}{@{}c@{}}{MNIST ($28\times 28$)} \\\cmidrule{4-5}%
model & n\_qubits & n\_features/1QNN & accuracy & loss  \\
\midrule
Distributed 14QNNs* & $7$ & $56$ & $0.96140$ & $1.51451$ \\
\botrule
\end{tabular*}
\end{table}

\section{Method}\label{sec:method}
\begin{figure}
    \centering
    \includegraphics[width=0.95\linewidth]{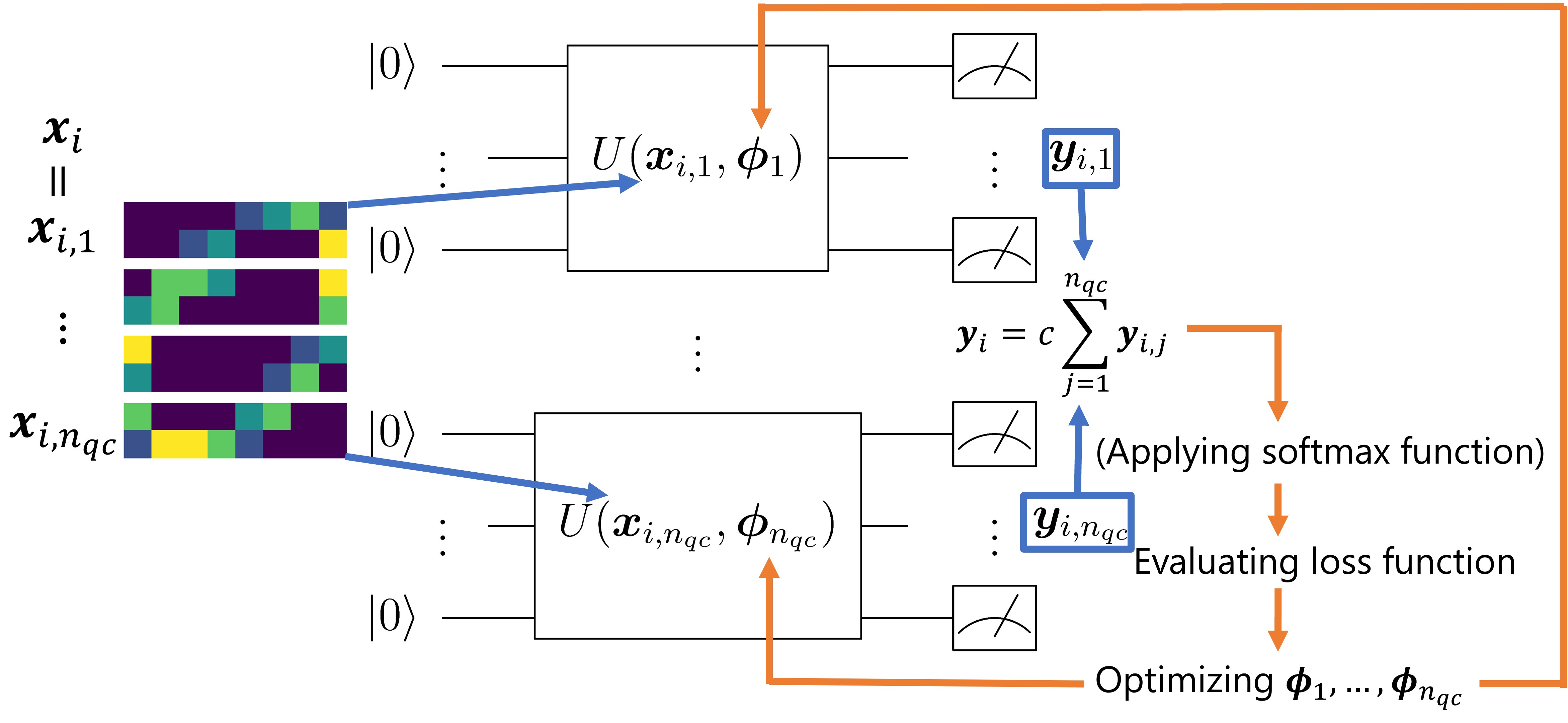}
    \caption{
    This figure shows the flow of our distributed approach over $n_\text{qc}$ QNNs. 
    First, we equally partition features $\bm{x}_i$ into $\{\bm{x}_{i,j}\}_{j=1}^{n_\text{qc}}$.
    We input partitioned features $\{\bm{x}_{i,j}\}_{j=1}^{n_\text{qc}}$ to variational quantum circuits $\{U(\bm{x}_{i,j},\bm{\phi}_j)\}_{j=1}^{n_\text{qc}}$. 
    Then, we evaluate a loss function using the sum of expectation values as outputs from the quantum circuits 
    and optimize the parameters $\{\bm{\phi}_j\}_{j=1}^{n_\text{qc}}$ in quantum circuits to minimize the loss function.}
    \label{fig:dist_qnns}
\end{figure}

In this section, we present our distributed QNNs approach, as shown in Fig.~\ref{fig:dist_qnns}. 
Our distributed QNNs model consists of $n_\text{qc}$ shallower and narrower quantum circuits $\{U(\bm{x}_{i,j},\bm{\phi}_j)\}_{j=1}^{n_\text{qc}}$, 
where each circuit processes a unique subset of the input features. 
These subsets $\{\bm{x}_{i,j}\}_{j=1}^{n_\text{qc}}$ represent the $j$th partition of the $i$th input data $\bm{x}_i$. 
We employ the expectation values with a set of observables $\{{O}^{(k)}\}_{k=1}^{d_\text{out}}$ for the outputs of the QNNs,  
where $d_\text{out}$ denotes the dimension of outputs. 
Here, we define the total outputs across all QNNs $\bm{y}_i$ corresponding to the input $\bm{x}_i$ as the sum of the expectation values: 
\begin{align} \label{eqn:qnn_exp}
\bm{y}_i 
= \Bigg ( &\sum_{j=1}^{n_\text{qc}} c \bra{0} 
U^\dagger(\bm{x}_{i,j},\bm{\phi}_j) O^{(1)} U(\bm{x}_{i,j},\bm{\phi}_j) \ket{0},\ldots, \nonumber \\
&\sum_{j=1}^{n_\text{qc}} c \bra{0} 
U^\dagger(\bm{x}_{i,j},\bm{\phi}_j) O^{(d_\text{out})} U(\bm{x}_{i,j},\bm{\phi}_j) \ket{0} \Bigg )
\end{align}
where 
$c$ is a constant value to adjust the outputs. 
For a classification task, we then apply the softmax function to normalize the outputs. 
The procedure of our model can be summarized as follows:
\begin{enumerate}
    \item Partitioning input feature $\bm{x}_i$ into $\{\bm{x}_{i,j}\}_{j=1}^{n_\text{qc}}$.
    \item Encoding the partitioned features $\{\bm{x}_{i,j}\}_{j=1}^{n_\text{qc}}$ into $n_\text{qc}$ QNNs respectively.
    \item Evaluating expectation values for each QNN with observables $\{O^{(k)}\}_{k=1}^{d_\text{out}}$.
    \item Calculating $\bm{y}_i$ in Eq.~\eqref{eqn:qnn_exp} 
    by summing the expectation values from each QNN and multiplying a constant value $c$.
    \item (For classification task, applying softmax function.)
    \item Calculating a loss function using $\{\bm{y}_i\}_{i=1}^N$ for regression task or using $\{\mbox{Softmax}(\bm{y}_i)\}_{i=1}^N$ for classification task, 
    where $N$ is the number of data and $\mbox{Softmax}(\cdot)$ is the softmax function.
    \item Optimizing parameters $\{\bm{\phi}_j\}_{j=1}^{n_\text{qc}}$ to minimize the loss.
\end{enumerate}
We used this procedure in our numerical experiments.

\section{Conclusion}\label{sec:concl}
In this paper, we have proposed a novel distributed QNNs approach 
encoding partitioned features across multiple shallower and narrower QNNs compared to a single large QNN. 
By using the sum of the expectation values from these independent QNNs, 
our distributed QNNs achieve superior performance compared to a single QNN.
However, our results imply that an excessive number of partitions reduces the performance. 
Nevertheless, we achieved high accuracy in classifying a large dataset of MNIST, 
which is a challenging task for classical simulations using a single large QNN.
Importantly, our distributed QNNs approach provides practical advantages that are compatible with current quantum devices. 
Specifically, our distributed QNNs approach effectively reduces qubit requirements and circuit depth for each individual QNN, 
and the shallower and narrower circuits may help in mitigating the vanishing gradient problems during optimization compared to a single large QNN. 

In our future work, we will enhance our approach 
by incorporating quantum communication between QNNs
to explore a quantum advantage in distributed quantum machine learning. 
We are also interested in ensembling the outputs of multiple quantum circuits 
encoding various different partitioned features, which may further improve performance, 
and encoding multichannel images, which could broaden the applicability of distributed QNNs. 
Overall, our results highlight the promise of distributed quantum algorithms to mitigate hardware restrictions and 
also pave the way for realizing the vast possibilities inherent in near-term quantum machine learning applications.

\backmatter

\bmhead{Funding}
This work is supported by JSPS KAKENHI Grant Number JP23K19954.

\bmhead{Data and Code Availabitilty}
Our jupyter notebooks with our results are available on GitHub (https://github.com/puyokw/DistributedQNNs).


\begin{appendices}

\section{The Detailed Model Description Used in Numerical Experiments} \label{secA:model_description}
Here, we describe the details of our QNN and distributed QNNs used in Sec.~\ref{sec:results}. 
As we mentioned in Sec.~\ref{sec:results}, 
the architecture difference between a single QNN and multiple QNNs 
is the number of qubits and encoding layers depending on the number of features. 
Below, we provide a comprehensive description of the QNN architectures.

\begin{figure}
    \centering
    \includegraphics[width=0.95\linewidth]{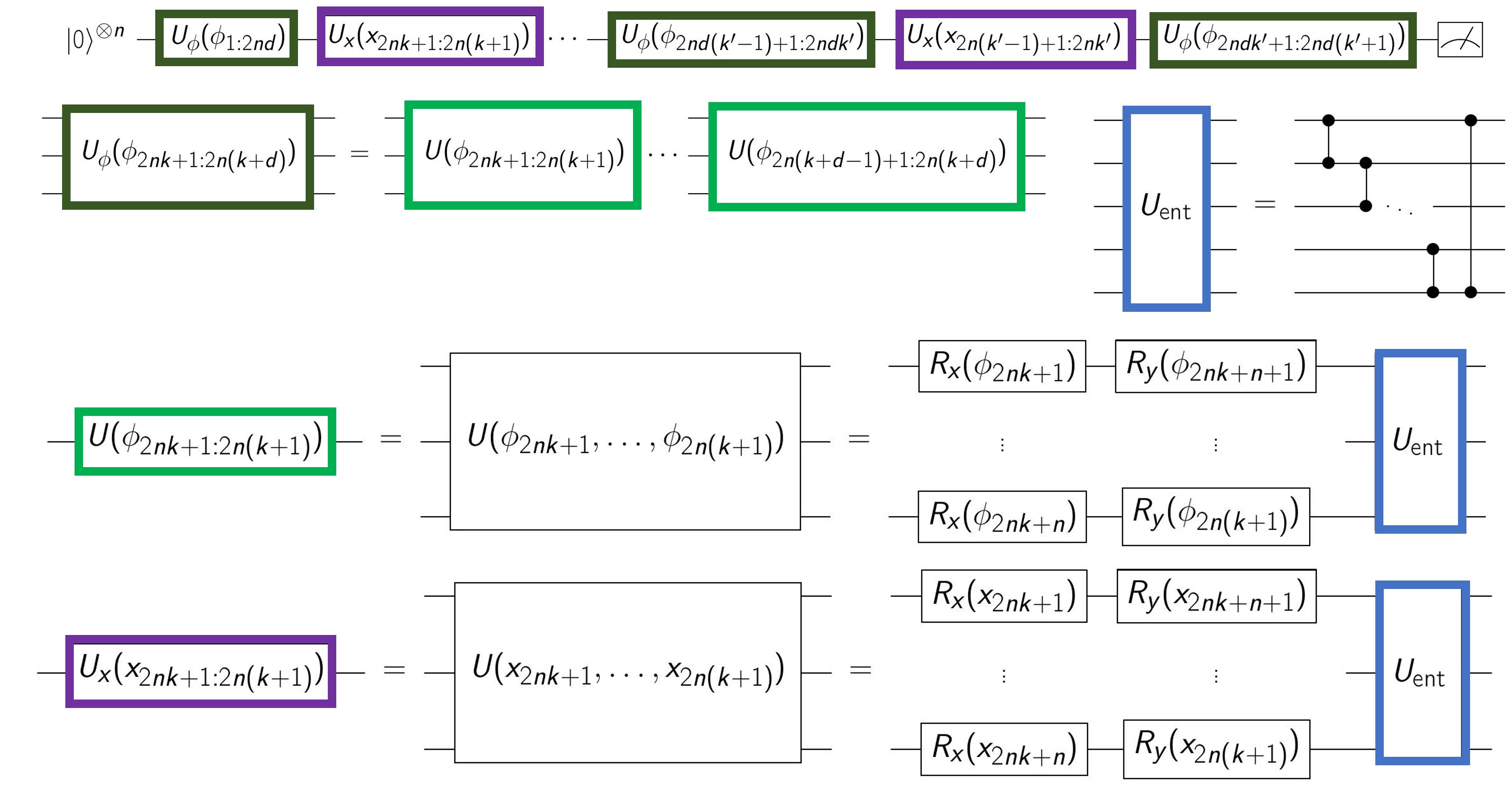}
    \caption{
    This figure shows our QNN architecture. 
    The detail is described in Appendix~\ref{secA:model_description}. 
    }
    \label{fig:qc}
\end{figure}

Our QNN architecture (Fig.~\ref{fig:qc}) consists of single qubit rotation gates RX and RY, and CZ gates. 
Our QNN architecture is characterized by alternating layers of a parameterized unitary transformation $U_\phi$ and a data encoding transformation $U_x$.

The data encoding transformation $U_x$ encodes the features 
through the angles of $n$ RX and $n$ RY gates acting on the $i$th qubit, where $n$ is the number of qubits. 
Then, $n$ CZ gates act on the $i$th target and $(i \mod n)+1$th control qubits. 
The total number of $U_x$ is $\lceil \mbox{n\_features}/2 \rceil$, where $\mbox{n\_features}$ denotes the number of features for a single QNN or partitioned features for each QNN. 

The unitary transformation $U_\phi$ consists of $20$ layers of unitary transformation $U$, 
consisting of $n$ RX gates and $n$ RY gates acting on the $i$th qubit, 
and $n$ CZ gates with the $i$th qubit as the control and the $(i \mod n)+1$th qubit as the target. 
The total number of $U_\phi$ is $\lceil \mbox{n\_features}/2 \rceil+1$.
Note that we excluded the CZ transformations $U_\text{ent}$ just before the measurement. 
In addition, the parameters in $U_\phi$ are initialized with uniform random values between $0$ and $\pi$.

\begin{figure}
    \centering
    \includegraphics[width=0.95\linewidth]{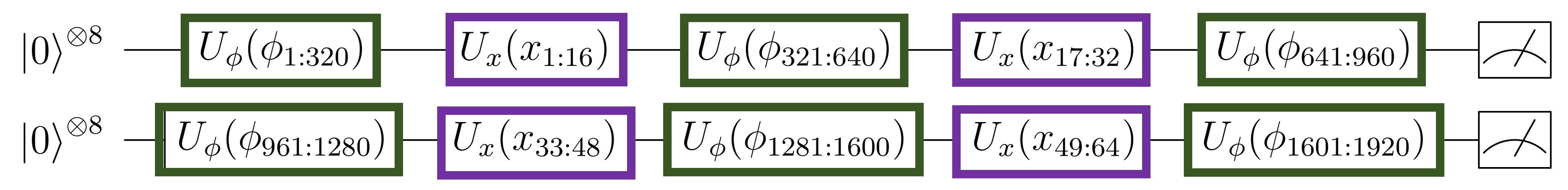}
    \caption{
    This figure shows the example of our QNN architecture used for classifying the $8 \times 8$ reduced sized Semeion dataset with two QNNs. Note that we excluded the CZ transformations $U_\text{ent}$ just before the measurement, as we mentioned in Appendix~\ref{secA:model_description}. 
    }
    \label{fig:2qc_semeion}
\end{figure}

For example, we describe the quantum circuit we used for classifying $8\times8$ sized reduced Semeion dataset with two QNNs, 
as shown in Fig.~\ref{fig:2qc_semeion}. 
As we mentioned above, 
we apply unitary transformations $U_\phi(\phi_{1:320})$ to $\ket{0}^{\otimes 8}$, $U_x(x_{1:16})$ for input first $16$ features, $U_\phi(\phi_{321:640})$ for transformation and entanglement, $U_x(x_{17:32})$ for input the following $16$ features, and $U_\phi(\phi_{641:960})$ excluding the last CZ transformations $U_\text{ent}$ just before the measurements. 
Then, we measure the expectation values $\bra{\psi_f}X_1\ket{\psi_f},\ldots,\bra{\psi_f}X_5\ket{\psi_f},\bra{\psi_f}Z_1\ket{\psi_f},\ldots,\bra{\psi_f}Z_5\ket{\psi_f}$, 
where $\ket{\psi_f}$ represents the quantum states just before measurements. 
Another QNN has the same architecture with independent parameters and inputs the last 32 features.

\end{appendices}


\bibliography{sn-bibliography}

\end{document}